\documentclass[preprint,prd,aps,nofootinbib,amssymb]{revtex4}
\usepackage{graphicx}
\usepackage{epsfig,amssymb,amsmath}

\newcommand{\beq}{\begin{equation}}
\newcommand{\eeq}{\end{equation}}
\newcommand{\bea}{\begin{eqnarray}}
\newcommand{\eea}{\end{eqnarray}}
\newcommand{\bear}{\begin{array}}
\newcommand {\eear}{\end{array}}
\newcommand{\bef}{\begin{figure}}
\newcommand {\eef}{\end{figure}}
\newcommand{\bec}{\begin{center}}
\newcommand {\eec}{\end{center}}
\newcommand{\non}{\nonumber}

\newcommand{\la}{\left\langle}
\newcommand{\ra}{\right\rangle}

\def\GEV#1{10^{#1}{\rm\,GeV}}

\def\lrf#1#2{ \left(\frac{#1}{#2}\right)}
\def\lrfp#1#2#3{ \left(\frac{#1}{#2} \right)^{#3}}

\begin{document}
\draft
\tighten
\preprint{DESY 14-017, TU-953, IPMU14-0041}
\title{\large \bf
7 keV sterile neutrino dark matter from split flavor mechanism
}
\author{
   Hiroyuki Ishida\,$^a$\footnote{email: h\textunderscore ishida@tuhep.phys.tohoku.ac.jp},
   Kwang Sik Jeong\,$^b$\footnote{email: kwangsik.jeong@desy.de},
   Fuminobu Takahashi\,$^{a,c}$\footnote{email: fumi@tuhep.phys.tohoku.ac.jp}
    }
\affiliation{
 $^a$ Department of Physics, Tohoku University, Sendai 980-8578, Japan\\
 $^b$ Deutsches Elektronen Synchrotron DESY, Notkestrasse 85,
         22607 Hamburg, Germany \\
 $^c$ Kavli IPMU, TODIAS, University of Tokyo, Kashiwa 277-8583, Japan
    }

\vspace{2cm}

\begin{abstract}

The recently discovered X-ray line at about $3.5$~keV can be explained by sterile neutrino
dark matter with mass, $m_s \simeq 7$~keV, and the mixing, $\sin^2 2\theta \sim 10^{-10}$.
Such sterile neutrino is more long-lived than estimated based on the seesaw formula,
which strongly suggests an extra flavor structure in the seesaw sector.
We show that one can explain both the small mass and the longevity based on the split flavor
mechanism where the breaking of flavor symmetry is tied to the breaking of the $B-L$ symmetry.
In a supersymmetric case we find that the $7$~keV sterile neutrino implies the gravitino
mass about $100$~TeV.

\end{abstract}

\pacs{}
\maketitle

\section{Introduction}

Roughly a quarter of the Universe consists of dark matter.
In spite of extensive dark matter searches conducted so far, its nature remains unknown.
If dark matter is made of as-yet-unknown species of particles, they must be cold and very long-lived.
The required longevity, however, does not guarantee the absolute stability of dark matter;
it may decay into the standard model (SM) particles, enabling us to probe the nature of dark matter
through indirect dark matter search.

Recently an unidentified X-ray line at about $3.5$~keV in the XMM-Newton X-ray observatory data of
various galaxy clusters and the Andromeda galaxy was reported independently by two
groups~\cite{Bulbul:2014sua,Boyarsky:2014jta}.
While there are a variety of systematic uncertainties that can affect the observed line energy and
flux, it is interesting that such X-ray line can be explained by sterile neutrino
dark matter~\cite{Dolgov:2000ew,Boyarsky:2009ix,Kusenko:2009up,Abazajian:2012ys,Drewes:2013gca,Merle:2013gea},
a long-sought dark matter candidate, with mass about $7$~keV.

The sterile neutrino dark matter with mass in the keV range is known to decay radiatively into
a photon and an active neutrino, producing a narrow X-ray line.
The lifetime depends on its mass and the mixing with active neutrinos.
For the mass of $7$~keV, the observed X-ray line can be explained by the mixing angle,
$\sin^2 2\theta \sim 7\times 10^{-11}$~\cite{Bulbul:2014sua,Boyarsky:2014jta}, which is just below
the previously known X-ray bound.
Such small mass and mixing can be partially understood by the split seesaw mechanism~\cite{Kusenko:2010ik}
or a simple Froggatt-Nielsen (FN) type flavor model~\cite{Froggatt:1978nt}.
In these scenarios, the seesaw formula~\cite{seesaw} remains intact even in the presence of large
mass hierarchy in the right-handed (sterile) neutrinos:
\bea
\left(m_\nu\right)_{\alpha \beta} = \lambda_{I \alpha}\lambda_{I\beta} \frac{v^2 }{M_{I}},
\label{seesaw}
\eea
where $v \equiv \la H^0\ra \simeq174$~GeV is the vacuum expectation value (VEV) of
the Higgs field,
$\lambda_{I \alpha}$ denotes the Yukawa coupling of the right-handed neutrino $N_I$ with
the lepton doublet $L_\alpha$ and the Higgs field, and $M_I$ is the mass of the right-handed
neutrino $N_I$.
The point is that both the neutrino Yukawa coupling $\lambda_{I \alpha}$ and the right-handed
neutrino mass $M_I$ are suppressed by either a geometrical
factor or flavor charge in such a way that the suppression factors are cancelled out
in the above seesaw formula.
This is because the suppression mechanism is independent of the U(1)$_{B-L}$ breaking.
The observed X-ray flux (as well as the previously known X-ray bound), however, requires that
the sterile neutrino dark matter should be more long-lived than estimated based
on the above seesaw formula.\footnote{
It is known that, if the sterile neutrino comprises all the dark matter, its contribution
to the light neutrino mass should be negligible to satisfy the X-ray
bounds~\cite{Asaka:2005an,Boyarsky:2006jm}.
}
The observed X-ray line therefore strongly suggests an extra flavor structure in the seesaw sector.

Before the discovery of the $3.5$~keV X-ray line, the present authors showed in
Ref.~\cite{Ishida:2013mva} that both the small mass and the small mixing just below the X-ray bound
can be achieved in the split flavor mechanism; we introduce two $B-L$ Higgs fields, one of which is
charged under single discrete flavor symmetry.
The point is that the VEV of the $B-L$ Higgs leads to both breaking of
the U(1)$_{B-L}$ symmetry and the flavor symmetry.
In this letter we revisit the split flavor mechanism in light of the recent discovery of
the unidentified X-ray line at $3.5$~keV, and show that the observed X-ray line can be nicely explained
in the split flavor mechanism.
In particular, we examine carefully the model parameters by taking account of numerical
coefficients of order unity, while those numerical coefficients were set to be unity
in Ref.~\cite{Ishida:2013mva} for simplicity.
In a supersymmetric case we will study the implications for supersymmetry breaking scale and show
that the gravitino mass about $100$~TeV is favored by the observed X-ray line.

\section{$7$\,keV sterile neutrino in the split flavor mechanism}

The interactions relevant to the seesaw mechanism read
\bea
-{\cal L} = \lambda_{I\alpha} \bar N_I L_\alpha H + \frac{1}{2}\kappa_I M \bar N^c_I N_I + {\rm h.c.},
\eea
for $M$ being the $B-L$ breaking scale.
Here $N_I$ ($I=1,2,3)$, $L_\alpha$ ($\alpha = e, \mu, \tau$), and $H$ are the right-handed neutrino, lepton doublet, and
Higgs scalar, respectively.
The right-handed neutrino masses are given by $M_I=\kappa_I M$.

We are interested in the case where $N_1$ is much lighter than $N_i$ ($i=2,3$), and couples more weakly
to the lepton doublet and Higgs scalar than $N_i$.
To parameterize such hierarchical structures in the right-handed neutrino sector, let
us introduce the suppression factors:\footnote{
We use a slightly different notation from Ref.~\cite{Ishida:2013mva}
to examine more precisely the properties of the sterile neutrino dark matter.
}
\bea
\kappa_1 &=& x^2 \kappa,
\nonumber \\
|\lambda_{1\alpha}| &=& x_\alpha \lambda,
\eea
with $x_\alpha\lesssim x$ for $x$ and $x_\alpha$ much smaller than unity.
Here $\kappa$ and $\lambda$ are the typical values of $\kappa_i$ and the Yukawa
couplings $\lambda_{i\alpha}$, respectively.
The active neutrinos obtain tiny masses of the order,
\bea
m_{\rm seesaw} \equiv
\frac{\lambda^2}{\kappa} \frac{v^2}{M},
\eea
through the seesaw mechanism.
To generate phenomenologically viable neutrino masses, one needs
$m_{\rm seesaw}\sim 0.1$~eV, implying $M$ around $10^{15}$~GeV for $\kappa$ and $\lambda$
of order unity.

After electroweak symmetry breaking, $N_1$ mixes with the active neutrinos due to
the coupling $\lambda_{1\alpha}$.
The mixing angle is given by
\bea
\theta^2 &\equiv& \sum_\alpha \frac{|\lambda_{1\alpha}|^2 v^2}{M^2_1}
= \epsilon^2 \frac{m_{\rm seesaw}}{M_1},
\eea
where we have defined
\bea
\epsilon^2 \;\equiv\;  \frac{\sum_\alpha x^2_\alpha}{x^2}.
\eea
The decay mixing angle of the sterile neutrino dark matter is estimated to be
\bea
\sin^2 2\theta \simeq
0.6 \times 10^{-10}\,
\left(\frac{\epsilon}{10^{-3}} \right)^2 \left(\frac{m_{\rm seesaw}}{0.1\,{\rm eV}}\right)
\left(\frac{m_s}{7\,{\rm keV}}\right)^{-1},
\eea
where $m_s \simeq M_1$ denotes the mass of the sterile neutrino $N_1$.
Thus $\epsilon$ should be around $10^{-3}$ if the sterile neutrino dark matter is
responsible for the observed X-ray line around 3.5~keV.
One is however led to $x \sim x_\alpha$, i.e. $\epsilon\sim 1$, in the simple FN
model or the split seesaw mechanism.
This implies that either all the three components of $\lambda_{1\alpha}$ are less
than $10^{-3}$ of the expected,
or the sterile neutrino mass $M_1$ is heavier by a factor of $10^6$ than the expected,
or a combination of these two.
If $x_\alpha/x$ takes a value of order unity randomly as in the neutrino mass
anarchy~\cite{Hall:1999sn,Haba:2000be}, it would require a fine-tuning of order
$\epsilon^3 \sim 10^{-9}$.
We call this fine-tuning problem as the longevity problem~\cite{Ishida:2013mva}.

Taken at face value, the longevity problem strongly suggests an extra flavor structure
in the seesaw sector.
The split flavor mechanism~\cite{Ishida:2013mva} provides a natural way to explain both
$m_s\simeq 7$~keV and $\epsilon\sim 10^{-3}$ simultaneously.\footnote{
In Ref.~\cite{Ishida:2013mva} we adopted $m_s \sim 10$~keV and $\epsilon\sim 10^{-3}$
as reference values, which are surprisingly close to the observed ones.
}
Before going into the details of the model, let us briefly summarize how this is
achieved.
The suppression mechanism is implemented by extending the seesaw sector to include two
(or more) $B-L$ Higgs fields, one of which is charged under discrete flavor symmetry.
Most important, the breaking of flavor symmetry is tied to the breaking of the $B-L$
symmetry.
Then one is led to
\bea
\epsilon \sim \frac{M}{M_{Pl}} \sim 10^{-3} \lrf{M}{\GEV{15}},
\eea
in non-supersymmetric models.
Here $M_{Pl} \simeq 2.435\times 10^{18}$~GeV is the reduced Planck scale.
In what follows, we will take $M$ around $10^{15}$~GeV assuming that $\kappa_i$
and the Yukawa couplings $\lambda_{i\alpha}$ are order unity.\footnote{
It is straightforward to suppress $\kappa_{i \alpha}$ and $\lambda_{i \alpha}$
by assigning additional FN flavor charges on $N_i$.
Our results are not changed even in this case.
}
Then the above relation tells that the observed X-ray line can naturally be explained by
the sterile neutrino dark matter.
As will be shown later, $m_s\simeq 7$~keV is obtained for appropriate $B-L$ Higgs VEVs.

On the other hand, in supersymmetric models, one can consider discrete flavor symmetry
or discrete $R$ symmetry.
The sterile neutrino mass is given by
\bea
m_s \sim m_{3/2} \lrfp{M}{M_{Pl}}{3},
\eea
for both cases, with $m_{3/2}$ being the gravitino mass.
The $\epsilon$ parameter can be collectively written as
\bea
\epsilon \sim \lrfp{m_{3/2} M_{Pl}^3}{M^4}{\pm \frac{1}{2}} \frac{M}{M_{Pl}},
\eea
where the minus sign in the exponent applies when $\lambda_{1\alpha}$ mainly
receives supersymmetric contribution in the case of discrete $R$ symmetry,
and the plus sign is for the other cases.
These relations show that, for $m_{3/2}\sim 10^5$~GeV and $M\sim 10^{15}$~GeV, the sterile
neutrino has the right properties to be the origin of the X-ray line.

\subsection{Non-supersymmetric case}

The seesaw sector includes two $B-L$ Higgs fields $\Phi$ and $\Phi^\prime$, and three
right-handed neutrinos $N_I$.
Let us take the charge assignment,
\vskip 0.5cm
\begin{center}
\begin{tabular}{|c||c|c|c|c|c|c|}
\hline
                    & $\quad\Phi\quad$   & $\quad \Phi^\prime \quad$ & $\quad N_1 \quad$ & $\quad N_i \quad$
                    & $\quad L_\alpha \quad$  &  $\quad H \quad$ \\
\hline
\,\,U$(1)_{{\rm B}-{\rm L}}$\,  &   $2$     &   $-6$     & $-1$  & $-1$   & $-1$ &  0  \\
\hline
$Z_4$               &   0    &   $-1$     & $1$  &  0    & 0 & 0  \\
\hline
\end{tabular}
\end{center}
\vskip 0.3cm
Then the U$(1)_{B-L}$ and flavor symmetry constrain the seesaw sector interactions as
\bea
-\Delta{\cal L} =
\frac{1}{2}\kappa_i \Phi \bar N^c_i N_i + \lambda_i \bar N_i L H
+ \frac{1}{2} \tilde \kappa \frac{(\Phi^5 \Phi^{\prime 2})^*}{M^6_{Pl}} \bar N^c_1 N_1
+ \tilde \lambda_\alpha \frac{(\Phi^3\Phi^\prime)^*}{M^4_{Pl}} \bar N_1 L_\alpha H
+ {\rm h.c.},
\eea
where we have assumed that the cut-off scale of the model is the Planck scale.
For the coupling constants of order unity, the model gives
\bea
\epsilon \sim \frac{M}{M_{Pl}}.
\eea
Under the assumption that there is no additional structure in the coupling constants,
we take the couplings of $N_1$ to be
\bea
\label{natural-coupling}
\frac{|\tilde \lambda_\alpha|^2}{\tilde \kappa} = \frac{\lambda^2}{\kappa}.
\eea
Then it follows
\bea
m_s &\simeq&
4.8 \,{\rm keV}
\times
\tilde\kappa\, r^2
\left(\frac{M}{10^{15}{\rm GeV}}\right)^7,
\nonumber \\
\sin^2 2\theta &\simeq&
0.3\times 10^{-10}
\left(\frac{m_s}{7{\rm keV}}\right)^{-1}
\left(\frac{m_{\rm seesaw}}{0.1{\rm eV}}\right)
\left(\frac{M}{10^{15}{\rm GeV}}\right)^2,
\label{eq:non-susy}
\eea
taking $\langle \Phi \rangle=M$ and $\langle \Phi^\prime \rangle=r M$.
The $B-L$ Higgs fields would generally have VEVs of a similar size, giving $r\sim1$.
Thus the sterile neutrino dark matter has the right mass and mixing to
account for the observed X-ray line.
Note that the relation between the mass and the mixing angle is independent of $r$.

\begin{figure}[t]
\begin{center}
\begin{minipage}{16.4cm}
\centerline{
{\hspace*{0cm}\epsfig{figure=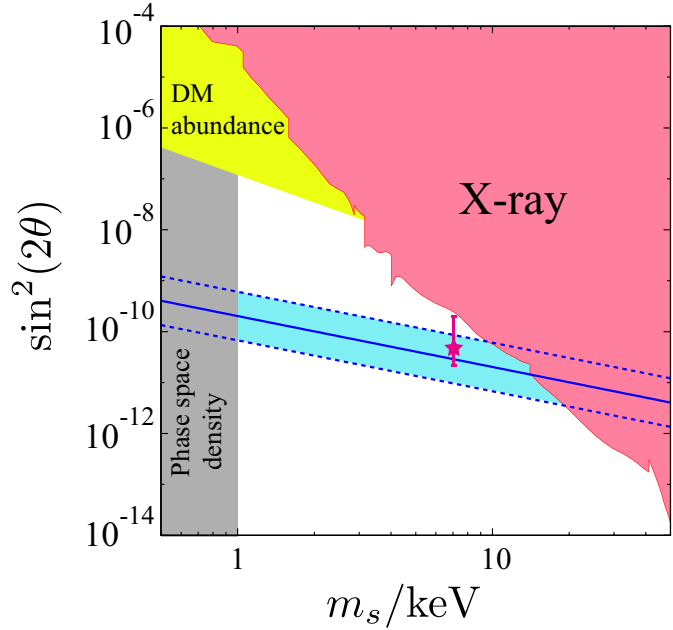,angle=0,width=9cm}}
}
\caption{
Properties of the sterile neutrino dark matter.
The solid (blue) line shows the relation (\ref{eq:non-susy}) between the mass and the mixing induced
by the split flavor mechanism in non-supersymmetric case.
The upper and lower dotted (blue) lines correspond to the cases of
three times larger and smaller values of $\sin^2 2\theta$, respectively. The red star represents
the values of the mass and the mixing that can explain the observed X-ray line. The shaded regions
denoted by X-ray and DM abundance are excluded by the X-ray observations~\cite{Abazajian:2012ys}
and the dark matter overproduction by the Dodelson-Widrow mechanism~\cite{Dodelson:1993je}.
We also show the region excluded by phase-space density~\cite{Boyarsky:2008ju}.
}
\label{fig:non-susy}
\end{minipage}
\end{center}
\end{figure}

Fig.~\ref{fig:non-susy} shows the relation (\ref{eq:non-susy}) between the mixing and the mass of
the sterile neutrino dark matter.
To take account of the uncertainty in the relation between couplings, we show as dotted (blue) lines
the results for cases with a numerical factor of $3$ and $1/3$ multiplied with
the RHS of Eq.~(\ref{natural-coupling}).
The shaded (light blue) region between the dotted lines represents the prediction of our model,
which satisfies various observational bounds.
We show the bounds from the non-detection of the X-ray line, the dark matter overproduction, and
phase-space density, together with the values suggested by the $3.5$~keV X-ray line.
We can see that our model can naturally explain the observed X-ray line at $3.5$~keV.

\subsection{Supersymmetric case}

The split flavor mechanism can be straightforwardly generalized to the supersymmetric case.
In contrast to the non-supersymmetric case, there are two important effects.
One is the holomorphic nature of the superpotential, and the other is the supersymmetry
breaking effects represented by the gravitino mass.

To cancel anomalies, we assign the $B-L$ charges to the left-handed chiral superfields as
\vskip 0.5cm
\begin{center}
\begin{tabular}{|c||c|c|c|c|c|c|}
\hline
                    & $\quad\Phi\quad$   & $\quad \Phi^\prime \quad$ & $\quad N_1 \quad$ & $\quad N_i \quad$
                    & $\quad L_\alpha \quad$ & $\quad H_u \quad$ \\
\hline
\,\,U$(1)_{{\rm B}-{\rm L}}$\,  &   $-2$                  &   $2$       & $1$  & $1$   & $-1$ & 0     \\
\hline
\end{tabular}
\end{center}
\vskip 0.3cm
where $H_u$ is the up-type Higgs doublet superfield.
The U$(1)_{B-L}$ is broken along the $D$-flat direction
\bea
|\Phi|^2=|\Phi^\prime|^2=M^2,
\eea
which is lifted by supersymmetry breaking effects associated with higher dimensional operators
or radiative corrections.

Small $\epsilon$ is obtained by imposing flavor symmetry under which $\Phi^\prime$ and
$N_1$ transform non-trivially.
Let us consider $Z_6$ flavor symmetry with\footnote{
Our results can be straightforwardly applied to the case of $Z_k$ with $k\geq 6$.
}
\vskip 0.5cm
\begin{center}
\begin{tabular}{|c||c|c|c|c|c|c|}
\hline
                    & $\quad\Phi\quad$   & $\quad \Phi^\prime \quad$ & $\quad N_1 \quad$ & $\quad N_i \quad$
                    & $\quad L_\alpha \quad$ & $\quad H_u \quad$ \\
\hline
$Z_6$               &   0                     &   1  &1 &  0   & 0 & 0     \\
\hline
\end{tabular}
\end{center}
\vskip 0.3cm
which allows the conventional superpotential terms for $N_i$
\bea
\label{W-seesaw}
W = \frac{1}{2}\Phi N_i N_i + N_i L_\alpha H_u,
\eea
while the relevant interactions of $N_1$  come from the following K\"ahler potential
\bea
\label{K-N1}
\Delta K =
\frac{\Phi^{\prime *}}{M_{Pl}}N_1 N_i
+ \frac{1}{2} \frac{(\Phi\Phi^{\prime 2})^*}{M^3_{Pl}} N_1 N_1 + {\rm h.c.},
\eea
in which we have dropped coupling constants of order unity.
For $M\gg m_{3/2}$, the effective action for the seesaw sector is obtained by
replacing the $B-L$ Higgs fields by their VEVs.
From $\Delta K$, one obtains
\bea
\Delta W_{\rm eff} =
\frac{1}{2}\tilde \kappa \frac{m_{3/2}M^3}{M^3_{Pl}} N_1 N_1
+ \tilde \lambda_\alpha \frac{m_{3/2}}{M_{Pl}} N_1 L_\alpha H_u,
\eea
for $m_{3/2}M_{Pl}\ll M^2$,
after redefining the heavy right-handed neutrinos to remove the mixing terms with $N_1$.
The above implies
\bea
\epsilon \sim \frac{\sqrt{m_{3/2} M_{Pl}}}{M}.
\eea
The detailed properties of the sterile neutrino dark matter depend on the
supersymmetry breaking scale.
Taking the coupling constants to be (\ref{natural-coupling}), we find
\bea
m_s &\simeq& 6.9\,{\rm keV}\times
\tilde \kappa \left(\frac{m_{3/2}}{10^5{\rm GeV}}\right)
\left(\frac{M}{10^{15}{\rm GeV}}\right)^3,
\nonumber \\
\sin^2 2\theta &\simeq&
0.4\times 10^{-10}
\left(\frac{m_s}{7{\rm keV}}\right)^{-1}
\left(\frac{m_{\rm seesaw}}{0.1{\rm eV}}\right)
\left(\frac{m_{3/2}}{10^5{\rm GeV}}\right)
\left(\frac{M}{10^{15}{\rm GeV}}\right)^{-2}.
\eea
Hence the sterile neutrino can explain the X-ray line for $m_{3/2}$ around $10^5$~GeV
and $M$ around $10^{15}$~GeV.

The suppression mechanism is successfully implemented also by a discrete
$R$ symmetry.
There are many different ways to assign the discrete $R$ charges.
Here we consider a simple case with\footnote{
Even though the $R$-parity is broken in the case of $Z_{5R}$ symmetry, the lightest supersymmetric
particle (LSP) remains stable due to the residual $Z_{2B-L}$ symmetry~\cite{Ishida:2013mva}.
Its thermal relic abundance can be suppressed for the Wino-like LSP, or in the presence
of late-time entropy production.
}
\vskip 0.5cm
\begin{center}
\begin{tabular}{|c||c|c|c|c|c|c|}
\hline
                    & $\quad\Phi\quad$   & $\quad \Phi^\prime \quad$ & $\quad N_1 \quad$ & $\quad N_i \quad$
                    &$\quad L_\alpha \quad$ & $\quad H_u\quad $\\
\hline
$Z_{5R}$               &   0                    &   $4$     & $3$  &  1      &1&0    \\
\hline
\end{tabular}
\end{center}
\vskip 0.3cm
Then the interactions of $N_i$ are again given by (\ref{W-seesaw}).
The relevant interactions of $N_1$ come from the following K\"ahler and super-potentials
after $B-L$ breaking,
\bea
\Delta K &=& \frac{\Phi^{\prime *}}{M_{Pl}}N_1 N_i + \frac{\Phi^2 \Phi'}{M_{Pl}^3} N_1 N_1
+ {\rm h.c.},
\non\\
\Delta W &=& \frac{(\Phi\Phi^\prime)^2}{M^4_{Pl}}N_1 L_\alpha H_u,
\eea
where order unity coefficients have been omitted.
At energy scales below $M$,
one finds the effective superpotential
\bea
\Delta W_{\rm eff} = \frac{1}{2}\tilde \kappa \frac{m_{3/2} M^3}{M^3_{Pl}} N_1N_1
+ \tilde \lambda_\alpha
\left( s \frac{m_{3/2}}{M_{Pl}} +\frac{M^4}{M^4_{Pl}} \right) N_1 L_\alpha H_u,
\label{W-for-discreteR}
\eea
for $m_{3/2}M_{Pl}\ll M^2$.
Here the constant $s$ is generally order unity.
If the neutrino Yukawa coupling of $N_1$ to $L_\alpha H_u$ is dominated by the supersymmetry
breaking contribution, the properties of the sterile neutrino are same as in the previous case.
So, we here focus on the case that the effective Yukawa coupling is dominated by
the supersymmetric contribution $\sim (M/M_{Pl})^4$.
This is the case for $s$ around or slightly smaller than unity in the parameter region with
$m_{3/2}\sim 10^5$~GeV and $M\sim 10^{15}$~GeV.
From the effective superpotential, $\epsilon$ is then estimated to be
\bea
\epsilon \sim \left(\frac{m_{3/2}}{M_{Pl}}\right)^{-1/2}
\left(\frac{M}{M_{Pl}}\right)^3,
\eea
for the coupling constants of order unity.
Let us take $\tilde\kappa$ and $\tilde \lambda_\alpha$ to be (\ref{natural-coupling})
assuming no extra structure.
Then the mass and mixing angle of the sterile neutrino dark matter are given by
\bea
m_s &\simeq& 6.9\,{\rm keV}\times \tilde \kappa
\left(\frac{m_{3/2}}{10^5{\rm GeV}}\right)
\left(\frac{M}{10^{15}{\rm GeV}}\right)^3,
\nonumber \\
\sin^2 2\theta &\simeq&
0.2\times 10^{-10}
\left(\frac{m_s}{7{\rm keV}}\right)^{-1}
\left(\frac{m_{\rm seesaw}}{0.1{\rm eV}}\right)
\left(\frac{m_{3/2}}{10^5{\rm GeV}}\right)^{-1}
\left(\frac{M}{10^{15}{\rm GeV}}\right)^6.
\eea
Note that the sterile neutrino mass has the same dependence on $m_{3/2}$ and $M$ as in the
case of $Z_6$ flavor symmetry, while the mixing angle has different dependence.

In both cases considered above, the gravitino mass close to $100$~TeV is favored by the observed
X-ray line.
Such heavy gravitino mass is consistent with the SM-like Higgs boson of mass near
126\,GeV~\cite{Aad:2012tfa,Chatrchyan:2012ufa}.

\section{Discussion and conclusions}

So far we have focused on the small mass and mixing of the sterile neutrino dark matter
suggested by the recently discovered $3.5$~keV X-ray line, and have shown that these properties
can be naturally realized in the split flavor mechanism.
In order to explain the observations, we need to generate a right amount of the sterile neutrinos.
Thermal production through mixings with active neutrinos, however, is inefficient for such small
mixing angle, unless there is a large lepton asymmetry.
Another possibility is through thermal production through U(1)$_{B-L}$ gauge interactions
at high temperature~\cite{Khalil:2008kp, Kusenko:2009up,Ishida:2013mva}; the abundance of the sterile
neutrino is given by
\bea
\Omega_{\rm DM} h^2 &\sim& 0.2
\lrfp{g_*}{106.75}{\frac{3}{2}}  \lrf{m_s}{7{\rm\,keV}}\lrfp{M}{\GEV{15}}{-4}
\lrfp{T_R}{5 \times \GEV{13}}{3},
\eea
where $T_R$ is the reheating temperature of the Universe after inflation.
At such high reheating temperature, a right amount of the baryon asymmetry can be created through
thermal leptogenesis~\cite{Fukugita:1986hr} by the two heavy right-handed
neutrinos~\cite{Endoh:2002wm,Raidal:2002xf}.
Non-thermal production may also work; see e.g. Ref.~\cite{Petraki:2007gq}.
Alternatively, if the $B-L$ symmetry is restored after inflation, the sterile neutrino $N_1$ will
be thermalized through the $B-L$ gauge interactions.
If there is a late-time entropy production of ${\cal O}(10^2)$, its thermal abundance can be
reduced so that it can explain the dark matter abundance~\cite{Kusenko:2010ik}.
In this case we need to introduce small explicit breaking of the discrete symmetry to make domain
walls annihilate before dominating the Universe.

In this letter we have revisited the split flavor mechanism in light of the recent discovery of the X-ray
line at $3.5$~keV in the XMM-Newton X-ray observatory data of various galaxy clusters and
the Andromeda galaxy.
In particular, the required small mixing angle, $\sin^2 2\theta \sim 7\times 10^{-11}$,
implies that the sterile neutrino dark matter is more long-lived than estimated based on
the seesaw formula, which strongly suggests an extra flavor structure in the seesaw sector.
Note that the seesaw formula holds in the split seesaw mechanism or the simple FN flavor model.
In the split flavor mechanism we introduce two $B-L$ Higgs fields, one of which is charged
under discrete flavor symmetry. Most important, the breaking of flavor symmetry is tied to the breaking of the $B-L$
symmetry.
We have shown in both non-supersymmetric and supersymmetric scenarios that the $7$~keV
sterile neutrino with mixing $\sin^2 2\theta \sim 7\times 10^{-11}$ can be realized easily.
The suppression of the mixing angle, namely, the longevity of the sterile neutrino dark matter
with respect to the expectation based on the seesaw formula, is due to the mild hierarchy
between the U(1)$_{B-L}$ breaking scale and the Planck scale.
In the supersymmetric scenarios, the small mixing is partially due to the smallness of
the gravitino mass compared to the Planck scale; the gravitino mass around $100$~TeV is
favored by the observed X-ray line, which is consistent with the SM-like Higgs boson
with $126$~GeV mass.

\section*{Acknowledgment}
\vspace{-0.5cm}

This work was supported by Scientific Research on Innovative
Areas (No.24111702 [FT], No. 21111006 [FT] , and No.23104008 [FT]), Scientific Research (A)
(No. 22244030 and No.21244033) [FT], and JSPS Grant-in-Aid for Young Scientists (B)
(No. 24740135) [FT], and Inoue Foundation for Science [HI and FT].
This work was also supported by World Premier International Center Initiative
(WPI Program), MEXT, Japan [FT].

\end{document}